\documentclass[sigconf,nonacm]{acmart}
\usepackage{tabularx}
\newcolumntype{g}{X}
\newcolumntype{s}{>{\hsize=.5\hsize}X}

\usepackage{graphicx}
\usepackage{caption}
\usepackage{subcaption}
\usepackage{hyperref}
\usepackage[compact]{titlesec}
\usepackage{url}
\usepackage{float}
\setcopyright{None}
\copyrightyear{2021}
\acmYear{2021}
\setcopyright{acmcopyright}\acmConference[WI-IAT '21]{IEEE/WIC/ACM International Conference on Web Intelligence}{December 14--17, 2021}{ESSENDON, VIC, Australia}
\acmBooktitle{IEEE/WIC/ACM International Conference on Web Intelligence (WI-IAT '21), December 14--17, 2021, ESSENDON, VIC, Australia}
\acmPrice{15.00}
\acmDOI{10.1145/3486622.3493959}
\acmISBN{978-1-4503-9115-3/21/12}

\begin{document}

\title{“I'll be back”: Examining Restored Accounts On Twitter}

\author{Arnav Kapoor$^*$}
\author{Rishi Raj Jain$^*$}
% \authornotemark[1]
\affiliation{
  \institution{IIIT Hyderabad}
%   \city{New Delhi}
  \country{}
}
\affiliation{
  \institution{IIIT Delhi}
%   \city{New Delhi}
  \country{}
}
\email{arnav.kapoor@research.iiit.ac.in }
\email{rishi18304@iiitd.ac.in}

\thanks{$^*$,$^\dagger$ denote equal contribution}

% \authornote{Authors contributed equally to this research.}
% \authornotemark[1]
% \affiliation{
%   \institution{IIIT Hyderabad}
%   \country{}
% }

\author{Avinash Prabhu$^\dagger$}
\author{Tanvi Karandikar$^\dagger$}

% \author{Ponnurangam Kumaraguru}
% \author{Ponnurangam Kumaraguru}

% \authornotemark[2]
\affiliation{
  \institution{IIIT Hyderabad}
  \country{}
}
% \authornotemark[2]\
\email{avinash.prabhu@students.iiit.ac.in}
\email{tanvi.karandikar@students.iiit.ac.in  }

\author{Ponnurangam Kumaraguru}
\affiliation{
  \institution{IIIT Hyderabad}
  \country{}
}
\email{pk.guru@iiit.ac.in}

\begin{abstract}
Online social networks like Twitter actively monitor their platform to identify accounts that go against their rules. Twitter enforces account level moderation, i.e. suspension of a Twitter account in severe cases of platform abuse. A point of note is that these suspensions are sometimes temporary and even incorrect. Twitter provides a redressal mechanism to `restore' suspended accounts. We refer to all suspended accounts who later have their suspension reversed as `restored accounts'. In this paper, we release the first-ever dataset and methodology \footnote{\url{https://github.com/rishi-raj-jain/WIIAT-Restored-Accounts-On-Twitter}} to identify restored accounts. We inspect account properties and tweets of these restored accounts to get key insights into the effects of suspension. We build a prediction model to classify an account into normal, suspended or restored. We use SHAP values to interpret this model and identify important features. SHAP (SHapley Additive exPlanations) is a method to explain individual predictions. We show that profile features like date of account creation and the ratio of retweets to total tweets are more important than content-based features like sentiment scores and Ekman emotion scores when it comes to  classification of an account as normal, suspended or restored. We investigate restored accounts further in the pre-suspension and post-restoration phases. We see that the number of tweets per account drop by 53.95\% in the post-restoration phase, signifying less `spammy' behaviour after reversal of suspension. However, there was no substantial difference in the content of the tweets posted in the pre-suspension and post-restoration phases.
\end{abstract}

% TODO: make sure you include. 

% \ccsdesc[500]{Information systems~World Wide Web}
% \begin{IEEEkeywords}
% Knowledge Graph; Uncertainly requirement Analysis; Multi-round dialogue; Cognitive Service Computing; chat-bots; Conversational AI Bot; Granular Computing.
% \end{IEEEkeywords}
% For peer review papers, you can put extra information on the cover
% page as needed:
% \ifCLASSOPTIONpeerreview
% \begin{center} \bfseries EDICS Category: 3-BBND \end{center}
% \fi
% For peerreview papers, this IEEEtran command inserts a page break and
% creates the second title. It will be ignored for other modes.
% \IEEEpeerreviewmaketitle
% \vspace{-2cm}

\maketitle
% \vspace{-0.2cm}
\section{Introduction}
\label{Introduction}
% \vspace{-0.05cm}

Over the years, social media platforms have emerged as an effective means to voice our opinions \cite{kapoor_advances_2018}. They have become a critical part of elections around the world \cite{zhuravskaya_political_2020}. Twitter in particular has been used both by the common public to engage in political discourse and by politicians to reach the voters \cite{twitter_literature}. Mass misuse of these platforms during recent political events like the 2020 U.S. Presidential Election
\footnote{https://blog.twitter.com/en\_us/topics/company/2020/2020-election-update}, including voter manipulation, spam and hate
\cite{noauthor_social_2019}, have led to multiple countermeasures by Twitter and other social media platforms. Twitter performs both tweet level moderation (removing a tweet for violating Twitter policy) and account level moderation (suspension of the account). The most recent high profile suspension of Donald Trump's Twitter account - @realDonaldTrump \footnote{https://blog.twitter.com/en\_us/topics/company/2020/suspension} raised discussions around Twitter's suspension policy.
 
\begin{figure}[H]
    \centering
     \includegraphics[width=\linewidth]{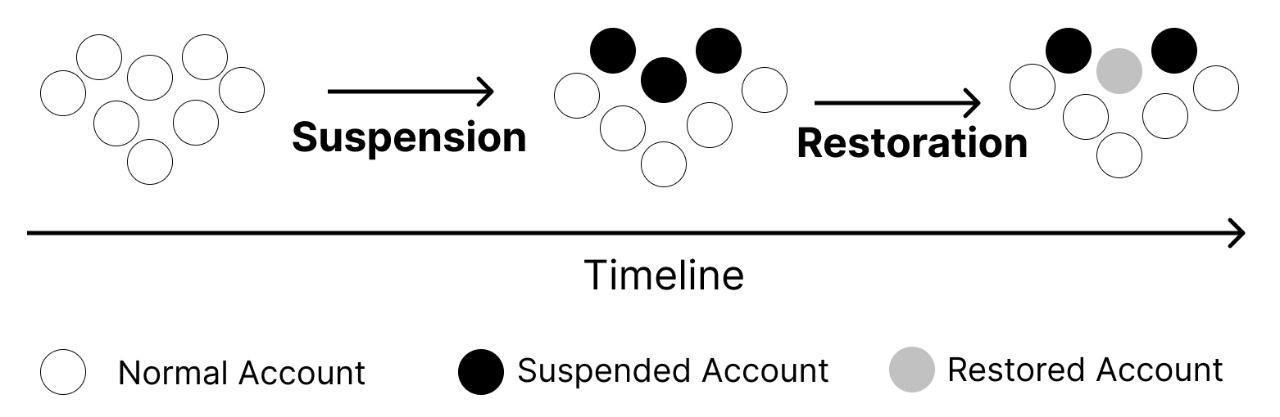}
     \caption{Restored accounts transition from normal to suspended state and then return to the normal state.}
     \label{Fig:Motivation}
\end{figure}

We note that some of these suspended accounts are unsuspended (or `restored') after a while and able use the platform again. Twitter policies give us an insight into the reasons for suspension as well as the means to restore the account. Common reasons for suspension by Twitter are spam, account at risk (hacked or compromised account) and abusive tweets. In some suspension cases, Twitter provides a means to restore the account by filing an appeal. 
In other cases, the suspension is temporary, and the account is automatically restored after a duration. The restored accounts allows us to understand the change in account activity in pre-suspension and post-restoration phases.. Henceforth, we use the term `restored' to refer to any such account initially in the suspended state and later returned to the normal state. Figure \ref{Fig:Motivation} shows the life-cycle of a restored account. 

Social media platforms can use this study on restored accounts to measure the change in behaviour pre-suspension and post-restoration to understand the effect of suspension. The transition of states from normal to suspended and back for a restored account raises questions about the efficacy of platform moderation. Platform moderation has to thread the fine line between infringement of freedom of speech, and enforcing platform policies and rules \cite{annur-polisci}. However, these platform moderation methods are black boxes without clear insights into what exact features or behaviour of an account can cause suspension and possible restoration. Hence, we use feature importance values on top of our classifier models to initiate discussion and form baselines to understand the factors that influence Twitter suspension. We do so by identifying the different aspects that characterise normal, suspended and restored accounts.

% \TK{ALmost all sentences in the following para start with "we"...}
This paper aims to model and understand the restored accounts and identify what differentiates them from the suspended and normal accounts. Normal accounts are the control group of active accounts which are neither suspended nor restored. We investigate different properties of all accounts 
% \AP{These properties are not only for restored right? They're for all classes} 
- the profile properties, the content tweeted and their interaction with other accounts on the Twitter platform. We then generate features to create a classifier to predict the category of an account. We use SHAP values to get feature importance scores to interpret prediction results. The aim is to better understand what properties set restored accounts apart from normal and suspended accounts. We then focus on the entire timeline of restored accounts (all tweets posted in 2019).
% \TK{This bracket implies that we weren't looking at all tweets before - which is true. But it is jarring to come to that realisation at this point in the paragraph. Maybe we should mention explicitly that SHAP values etc are coming from just election data}
We look at the activities and behaviour in two time phases - pre-suspension and post-restoration.
To the best of our knowledge, no previous work has focused on the category of restored accounts. We set out to answer the following Research Questions (RQs):

\begin{itemize}
    \item RQ1 - What factors differentiate the three categories of accounts - normal, suspended and restored?
    \item RQ2 - What are the changes, if any, in how an account interacts with the platform post-restoration?
\end{itemize}

For RQ1, we find that specific attributes of accounts like the creation date and frequency of tweets can act as excellent indicators to distinguish between categories, while other content features are not good differentiators. This higher importance of account properties over content features suggests that most account level suspensions are not due to content tweeted but rather the associated meta-information and activity of the account. Additionally, we find that restored accounts resemble normal accounts more than suspended accounts for the majority of explored features. 

For RQ2, we find that the number of tweets posted by the restored accounts in the post-restoration phase dropped by an average of 53.95\% as compared to the pre-suspension phase. However, the content of the tweets by restored accounts did not change drastically.

In adherence to the FAIR Principle \cite{wilkinson_fair_2016} and encouraging collaborations, we also make public the first-ever dataset on these restored accounts and provide a methodology to collect these accounts.

\section{Background and Related Work}

Suspension is one of the most common ways of platform moderation. Previous works studying suspended accounts \cite{thomas_suspended_2011,abu-el-rub_botcamp:_2019,chowdhury_twitter_2020}
 on Twitter have looked into multiple aspects of these accounts, primarily focusing on spam \cite{thomas_suspended_2011} and bot networks \cite{abu-el-rub_botcamp:_2019}.
We look at the Twitter policy for suspension to give us additional context regarding suspension on the platform.
Twitter's policy states that accounts are suspended in the following cases, \emph{`Most of the accounts we suspend are suspended because they are spammy, or just plain fake'}. The policy also says, if an account engages in abusive behaviour, like sending threats to others or impersonating other accounts, Twitter might suspend them. 
Twitter also gives information related to restoration of suspended accounts. In some cases, the suspension is temporary, and the account is restored to its original state after a duration. An account may also get suspended by mistake -- in this case, Twitter works with the account holder to restore the account.
The category of restored accounts on Twitter has not been studied in the past.

Past literature has also created models to predict suspended accounts \cite{volkova_identifying_2017} for credibility analysis. We build upon this work to focus on the interpretability of similar prediction models. The interpretable models allow us to identify why platform moderation algorithms consider certain accounts to be less credible and more likely to be suspended.   

Platform moderation and suspension has been studied in the context of other social media networks; primarily Reddit \cite{chandrasekharan_you_2017,ribeiro_does_2020}.
The focus is on community-level suspension, i.e. banning of a particular subreddit or a community of accounts, and the forced migration to either different platforms~\cite{ribeiro_does_2020} or different communities (subreddits)~\cite{chandrasekharan_you_2017}. Restored accounts on Twitter, however, do not migrate but rather return to the same state. The account returns to the same environment, which allows us to investigate the impact of suspension better as no external factors like change of community or platform is at play. These external factors in migration studies make it hard to identify if the change in account behaviour was due to the suspension event or simply a change of rules of the new community or platform. Our interpretability studies give insights into the cause of suspension and tell us about salient features of different account categories - normal, restored and suspended.

\section{Dataset}

\begin{figure*}
     \includegraphics[width=0.9\linewidth]{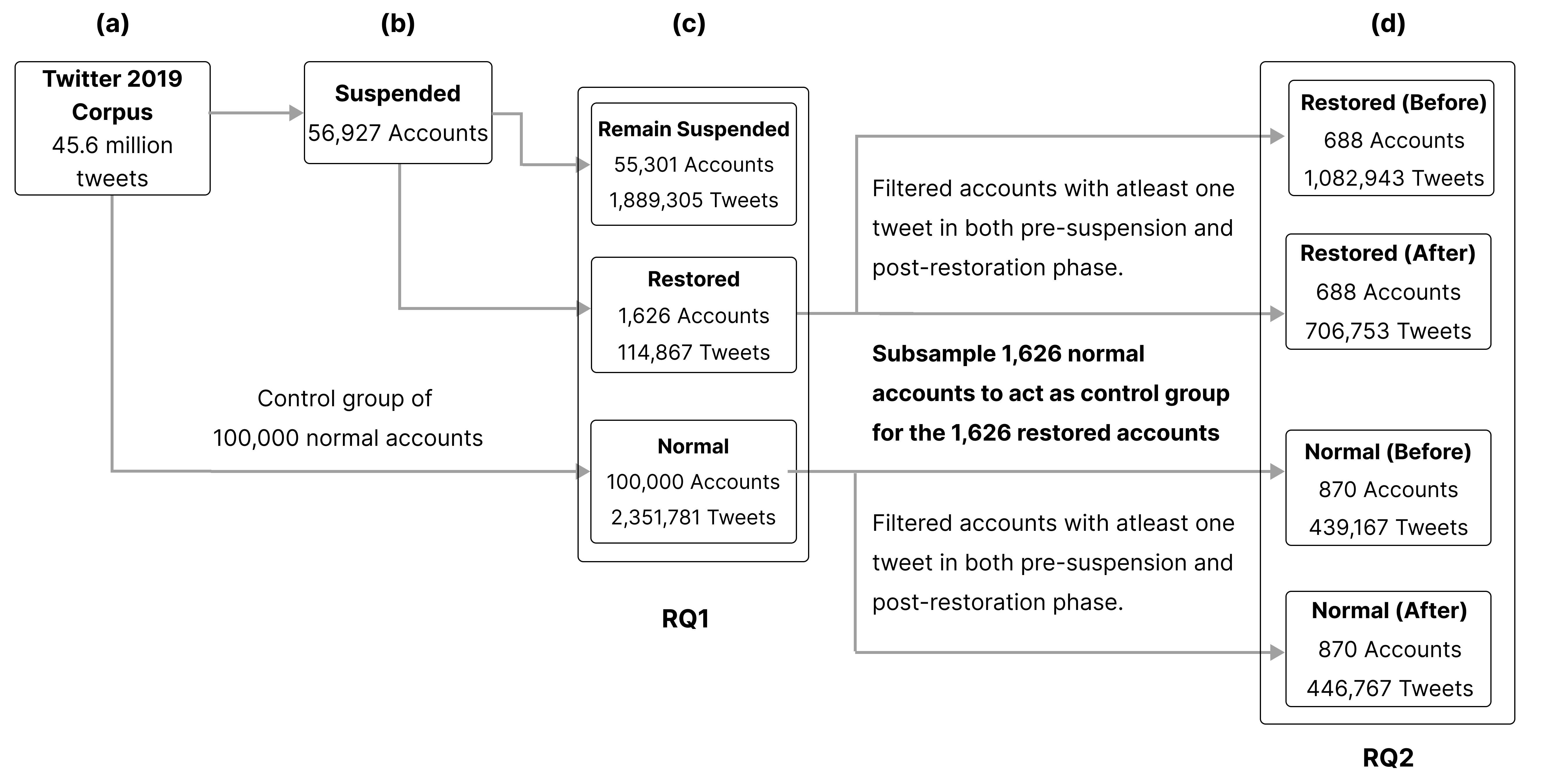}
    \caption{This flowchart depicts the different components of our research approach. We used the corpus of the suspended and normal accounts from the AGE2019 dataset.  
    We then queried all the suspended accounts, and found that 1,626 accounts were active again on the platform (referred to as restored accounts). We also fetch all the tweets made by the restored accounts in two phases: pre-suspension (1 Jan, 2019 - 29 Jun, 2019) and post-restoration (29 Jun, 2019 - 31 Dec, 2019). Further, 1,626 normal accounts were subsampled to act as the control group for the restored accounts. Tweets for this control group of normal accounts were also collected for the two phases. For restored and control group of normal accounts, we filtered the accounts with atleast one tweet in both pre-suspension and post-restoration phase.}
     \label{Fig:datapipeline}
\end{figure*}

In this section, we describe our data collection process to collect information regarding restored accounts on Twitter. The entire data collection pipeline is summarized in Figure \ref{Fig:datapipeline}.

% \subsection{RQ1 - Identify Categories of Accounts} \label{Identifying Categories of Users}
\textbf{RQ1 Dataset - Identify Categories of Accounts} To identify suspended and, in turn, restored accounts, we use the `Analysis of General Elections 2019 in India' (AGE2019) dataset  \cite{singh-a}. This dataset has been created by collating a large corpus of Twitter accounts that tweeted during the 2019 Indian general election. We specifically chose election data for our study since manipulation and misuse of the Twitter platform is rampant during an election, leading to a more significant proportion of suspended (and consequently restored) accounts \cite{le_postmortem_2019}.

The AGE2019 dataset has about 45.6 million tweets made by 2.2 million unique accounts over six months from February 5, 2019 (two months before first polling) to June 25, 2019 (one month after election results). 
Some of the accounts which tweeted during this phase were suspended by Twitter. The AGE2019 dataset also released a list of 56,927 accounts (2.8\% of the total 2.2 million) which were found to be suspended as of June 29, 2019. These were identified to be suspended as the Twitter API returns \emph{code 63} for suspended accounts. The AGE2019 dataset also comes with a list of 100,000 normal accounts (\textit{control group}) -- active accounts on Twitter as of June 29, 2019. 

To identify restored accounts we needed to find accounts that reverted to the normal state from the suspended state, hence we rechecked the status of the 56,927 suspended accounts. We queried the Twitter API on October 8, 2019 (around 3 months after June 29, 2019) for each of these accounts. Out of these 56,927 suspended accounts, 1,626 were active and restored to the normal state. The Twitter API no longer returned \emph{code 63} (or any other errors) for these 1,626 accounts. Table ~\ref{tab:data_stats} contains the summary of the different types of accounts used in our study.

\begin{table}[H]
\begin{center}
% \begin{adjustbox}{max width=\columnwidth}
% 	\centering
	\begin{tabularx}{\columnwidth}{lrr}
	    \hline
		\toprule
		\textbf{Category Of Accounts} &
		\textbf{Number of Accounts} &
		\textbf{\# of Tweets} \\
		\midrule
            Normal  &
            100,000 &
            2,351,781 \\ 
            Suspended  & 
            55,301 & 
            1,889,305  \\
            Restored  &
            1,626 &
            114,867 \\
		\bottomrule
	\end{tabularx}
% \end{adjustbox} 
\end{center}
% \vspace{0.2cm}
\caption{Statistics of normal, suspended and restored accounts identified in the context of Indian general elections 2019. (Used in RQ1)}
\label{tab:data_stats}
\end{table}

% \subsection{RQ2 - Timeline of Restored Accounts}
% \label{Timeline of Restored Users}

\begin{figure*}
   \begin{subfigure}{0.32\textwidth}
     \includegraphics[width=\linewidth]{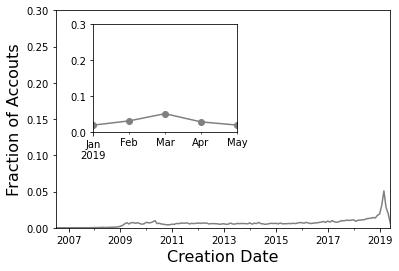}
     \caption{Normal Accounts}
     \label{Fig:standard_creation}
   \end{subfigure}\hfill
   \begin{subfigure}{0.32\textwidth}
     \centering
     \includegraphics[width=\linewidth]{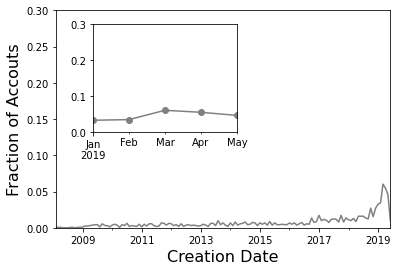}
     \caption{Restored Accounts}
     \label{Fig:restored_creation}
   \end{subfigure}\hfill
   \begin{subfigure}{0.32\textwidth}
     \raggedright
     \includegraphics[width=\linewidth]{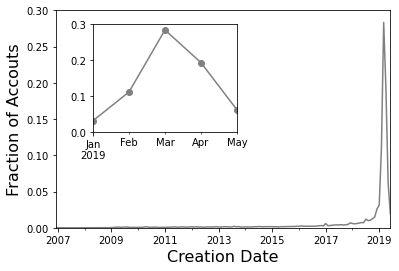}
     \caption{Suspended Accounts}
     \label{Fig:suspended_creation}
   \end{subfigure}
   \caption{Creation timeline of (a) normal accounts, (b) restored accounts, and (c) suspended accounts. Suspended accounts had a far larger proportion of accounts created in March-April 2019 as compared to the normal and restored accounts.}
%   \TK{Todo: }
   \label{fig:account_creation}
\end{figure*}

\textbf{RQ2 Dataset - Timeline of Restored Accounts} To get a more holistic view of the restored accounts, we collected tweets made by the restored accounts for the entirety of 2019 instead of just the political tweets in the election period. We crawled the entire timeline of these 1,626 restored accounts and collected a year's worth of data from Jan 1, 2019, to Dec 31, 2019, and got a total of about 1.78 million tweets made by these accounts. The complete timeline is a more unbiased representation of the accounts as it contains both election and non-election tweets.

We also crawled the entire timeline for 1,626 (equal to the number of restored accounts) randomly sampled normal accounts for the entirety of 2019. The normal accounts act as our control group while analysing restored accounts. We cannot collect the timelines for suspended accounts as their tweets are no longer available. We now have an extensive list of accounts and their corresponding tweets. We then split the timeline of restored accounts into two phases; pre-suspension (tweets before June 29, 2019) and post-restoration (tweets after June 29, 2019). We did this to quantify the impact of suspension and see how it affected account behaviour and its usage of the platform. We also split the control group of normal accounts along the exact dates. Additionally, we filtered out and kept only those accounts that had tweets in both the phases of pre-suspension and post-restoration.
% , which filtered down our initial set of accounts further, leading to 
The final set of accounts and tweets are summarised in Table ~\ref{tab:timeline_data_stats}. 

\begin{table}[h!]
    \centering
    \begin{tabularx}{\columnwidth}{gss}
        \hline
		\toprule
        \textbf{}     & Normal     & Restored     \\ \hline
        Number of Accounts         & 870        & 688         \\
        \# of Tweets Pre-Suspension         & 439,167        & 1,082,943         \\ 
        \# of Tweets Post-Restoration& 446,767        & 706,753         \\ \hline
    \end{tabularx}
	\caption{Statistics of normal and restored accounts for all tweets made in 2019. (Used in RQ2)}
	\label{tab:timeline_data_stats}
\end{table}

\section{RQ1 - Comparing Account Categories}
To address RQ1, we investigate the properties of restored accounts and identify features that distinguish between the three categories of accounts (normal, suspended and restored). We also conduct comparisons between these categories. To understand these categories of accounts comprehensively, we use three dimensions of analysis in our study. These are the account properties [\ref{Account Creation}],  the content tweeted [\ref{Content Characterstics}] and interaction with other accounts on Twitter {[\ref{Network Effects}}].
% \TK{Which network? Need to specify}.
We finally model features using the above properties and combine them to create a classifier to predict the account type. The focus while creating the classifier is on creating an interpretable model to shed some light on the relatively opaque moderation methodologies and identify features that distinguish normal, suspended and restored accounts. 
% \vspace{-0.12cm}
\subsection{Account Properties} \label{Account Creation}

One of the established aspects of suspended accounts is that they are very short-lived compared to normal accounts \cite{thomas_suspended_2011}, which led us to look at account creation dates.  
% hus, the account creation date was one aspect that we wanted to look at in detail and especially see creation date of restored users and where they stand compared to others.
We found a stark difference between the creation dates of suspended and normal accounts. We found that more than 47.60\% of the suspended accounts were created in the two months preceding the election (March and April 2019). In comparison, only 7.90\% of the normal accounts were created in this period (observe Figures \ref{Fig:standard_creation} and \ref{Fig:suspended_creation}).

% shows that more than 47.60\% of suspended accounts were created in the months of March and April 2019, 

% the two months preceding the election. In comparison, the number is far lower for the normal accounts with only 7.90\%.

Similar to the trend observed in normal accounts, only 11.50\% of restored accounts were created in March and April 2019. This shows that restored accounts resemble normal accounts when it comes to account creation patterns (observe Figures \ref{Fig:standard_creation} and \ref{Fig:restored_creation}). The sharp peak in March 2019 for suspended accounts suggests that there was a bulk creation of accounts before the election that ended up being suspended (observe Figure \ref{Fig:suspended_creation}). 

% \TK{suggest to remove this paragraph, as is confusing the reader. We have already established resemblance between normal and restored accounts} 

\begin{figure}[]
    \centering
     \includegraphics[width=\linewidth]{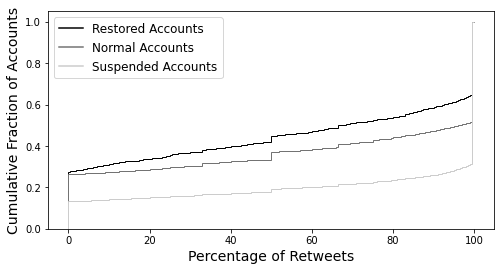}
     \caption{CDF plots of percentage of retweets. Suspended account retweet more than normal and restored accounts. 
    %  Suspended accounts retweet more than Normal and Restored accounts.
     }
     \label{fig_cdf_plots}
\end{figure}

% \AP{This paragraph is incomplete, no insights given, maybe bring back meta data?}
% \TK{Suggest to use this para as a one-line intro to section 4.1}
We now look at each account's proportion of retweets to tweets to see what an account does more-- retweeting or tweeting original content.
We plot cumulative distributions (Figure \ref{fig_cdf_plots}) of the percentage of retweets. Suspended accounts use retweets more than normal and restored accounts. About 70\% of all suspended accounts only retweet and have zero unique tweets. On the other hand, the percentage of accounts that only retweet are much lower for normal (48\%) and restored (35\%). Overall, we see a sharp distinction in the proportion of retweets, with suspended accounts using retweets far more frequently as opposed to normal and restored accounts. 

We further analyse other account properties such as no. of followers, no. of following, favorite count, etc. as part of our interpretable classifier in Section \ref{sec:classifier}. The exhaustive list of account properties analysed is listed in Table ~\ref{tab:features_used}. Most of these features had low predictive power as evident in \ref{fig:three graphs}.

% \subsection{User Meta Data} \label{User Meta Data}

\subsection{Content Characterstics}
\label{Content Characterstics}

\begin{figure}[]
  \begin{subfigure}[b]{\columnwidth}
  \centering
     \includegraphics[width=\linewidth]{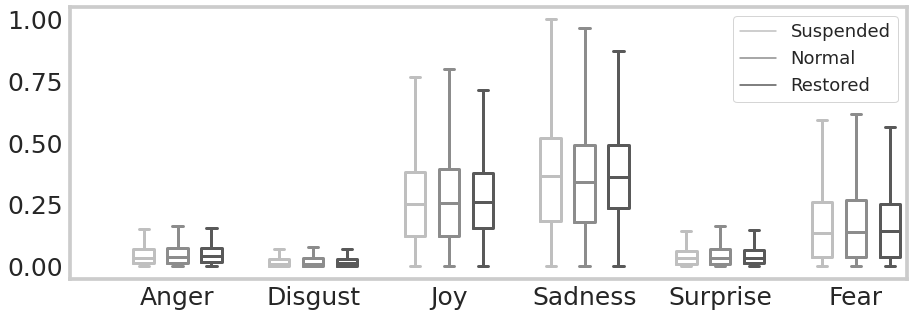}
     \caption{Ekman Scores}\label{Fig:ekman}
  \end{subfigure}
  \begin{subfigure}[b]{\columnwidth}
     \centering
     \includegraphics[width=\linewidth]{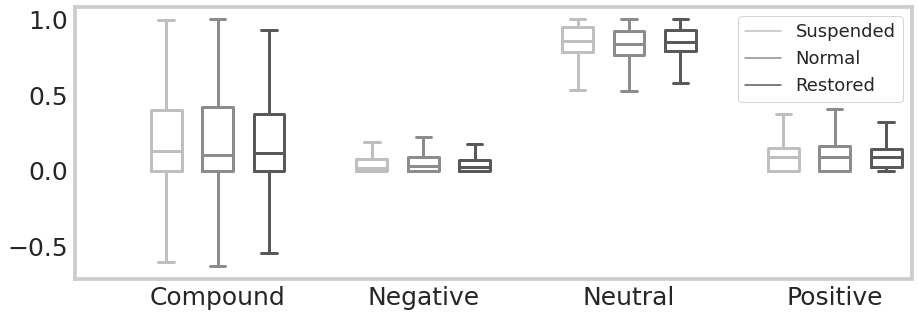}
     \caption{VADER Scores}\label{Fig:vader}
  \end{subfigure}
\caption{Box plots for Ekman and VADER scores for normal, restored and suspended accounts. Observe how all values almost remain the same across classes.}
  \label{Fig:box_plots_vader_ekman}
\end{figure}

In this section, we examine the content of the tweets made by the accounts. We apply language detection to filter and keep only English tweets for content analysis. We looked at two properties -- Ekman emotion scores and VADER sentiment scores. 
% Lexical diversity is the measure of the richness, i.e. quality of vocabulary of a corpus. 
We use these properties to further analyse the differences between the classes of accounts. Ekman emotions provide scores for six basic emotions \cite{ekman_argument_1992}. % \TK{Please rewrite the next sentence} 
We used emotion recognition models \cite{colneric_emotion_2020}
% linguistic tool and emotion recognition \AP{linguistic tool and emotion recognition is confusing} \cite{colneric_emotion_2020} 
to compute the Ekman scores depicted in Figure \ref{Fig:ekman}. We notice that across the emotions, there was no significant difference between the three groups - normal, suspended and restored. To verify this lack of significant difference, we also looked at the VADER sentiment scores for the three groups. VADER (Valence Aware Dictionary and sEntiment Reasoner) \cite{hutto_vader:_2014} is a rule-based sentiment analyser. The VADER Scores shown in Figure \ref{Fig:vader} also show no significant difference across the groups, indicating that the content characteristics were not a factor in distinguishing the category of an account.

% \vspace{-0.5cm}
\subsection{Network Effects} \label{Network Effects}
% \TK{This section is not at par with other sections.. story needs to be bit more clear etc}
In this section, we look at how suspended and restored accounts interact on the platform, and detect communities to identify group level characteristics. We use the Leiden algorithm to form the retweet and user mentions communities, and further observe how the interaction between accounts varies intra-community vs inter-community.
% \RRJ{Include Normal Accounts}
% \subsubsection{Detecting Communities} \label{Detecting Communities}
% To understand the impact of network activities on account suspension, we take steps to analyze and detect communities.

A community is a subset of nodes within the network such that connections between the nodes are denser than connections with the rest of the network \cite{radicchi_defining_2004}.
% \AP{cite}. % https://www.pnas.org/content/101/9/2658 
While several factors may show an account's involvement in a certain set of groups/activities, explicit user interaction is visible in retweets and user mentions. 
% As a user can both retweet or mention another account, we use these interactions to help gather insights into the account behaviour. 
An interaction is defined either as an account X mentioning account Y (mentions network), or an account X retweeting a tweet by account Y (retweet network). The nodes of such a network are accounts, and the directed edges represent retweets or user mentions. We used the Leiden algorithm \cite{traag_louvain_2019} to identify a set of well-connected communities that are guaranteed to be locally optimally assigned. 
% \AP{Didn't understand the last part of the previous sentence}
In general, the Leiden algorithm when applied, iteratively creates communities such that the intra-community connections are guaranteed to be more concentrated when compared to inter-community. To measure the strength of division of a network into communities, we use the modularity score.
% Skipping [cite] for now pretty common honestly
Networks with high modularity have dense connections between the nodes within communities but sparse connections between nodes in different communities. We construct two networks -- the user mention network and retweet network. These networks are constructed for all three categories of accounts.

In the retweet network, we observe a greater modularity for suspended accounts (0.746) as compared to normal (0.681) and restored accounts (0.638) (observe Table \ref{tab:retweet_data_stat}). This indicates a greater separation between the communities of suspended accounts. We also see a large number of suspended retweet communities being formed, hence denoting a larger spread of the information and possibly a means to engage more accounts into the agenda via retweets.

For the user mention network, we see that all categories of accounts have similar modularity (observe Table \ref{tab:user_mention_data_status}). Similar modularity indicates that the network structure is similar with respect to separation across communities and interaction within communities.

\begin{table}[]
	\centering
	\begin{tabular}{lrrr}
	    \hline
		\toprule
		\textbf{Retweets Network} & 
		\textbf{Normal} &
		\textbf{Suspended} & 
		\textbf{Restored} \\ 
		\midrule
		    Modularity & 
		    0.681 & 
		    0.746 & 
		    0.638 \\ 
            Number of Communities &
            7,190 & 
            11,748 & 
		    299 \\
            Largest Community Size &
            15.88 \% & 
            11.55 \% & 
		    15.96 \% \\
		\bottomrule
	\end{tabular}
	\caption{Retweets network statistics of normal, suspended and restored accounts. Largest community size for the suspended class is smaller than the largest community size for restored or normal class.} % 	The greater number of clusters in the suspended network, denote that the restored and normal communities are more cohesive than the suspended communities.
	\label{tab:retweet_data_stat}
\end{table}
    % \vspace{0.5cm}
\begin{table}[]
	\centering
	\begin{tabular}{lrrr}
	    \hline
		\toprule
		\textbf{User Mentions Network} & 
		\textbf{Normal} &
		\textbf{Suspended} & 
		\textbf{Restored} \\ 
		\midrule
		    Modularity & 
		    0.550 & 
		    0.575 & 
		    0.546 \\ 
            Number of Communities &
		    2,488 & 
            590 & 
		    117 \\
            Largest Community Size &
		    14.82 \% & 
            24.96 \% & 
		    17.38 \% \\
		\bottomrule
	\end{tabular}
	\caption{User mentions network statistics of normal, suspended and restored accounts. The modularity is comparable across categories of accounts. In contrast to retweet network (Table \ref{tab:retweet_data_stat}), suspended accounts have the largest community size in the user mention network.} %  This shows greater cohesiveness between suspended accounts compared to normal and restored accounts in user mentions network.
	\label{tab:user_mention_data_status}
\end{table}
% because the modularity of suspended user mention network and restored user mention network are comparable, while they differ significantly for the retweet network.

% Figure ~\ref{fig:retweets_network} shows the top 5 communities identified from the retweet graph. It is evident that there is comparatively much less interaction between the communities to the interaction visible within the communities. This is in contrast to what is observed from Fig. ~\ref{fig:mention_network}, which shows that there is much greater interaction between the user mentions communities w.r.t to the interaction of the users inside their communities. It is interesting to note that there are comparatively greater number of edges (3 times) between suspended to Normal involving user mentions as the mean of interaction, hence denoting a larger spread of the information and possibly a means to engage more users into the agenda via user mentions. It was observed that the top interactions that occurred via user mentions from these suspended accounts were owned by reputed election candidates, Indian election parties, and news channels.
% \vspace{-0.2cm}
\subsection{Classifier}

\label{sec:classifier}
We used an interpretable classifier to further understand the role of the three aspects -- account properties, content tweeted, and interactions within the network. We combine features from these aspects for training the classifier to see which features play a crucial role in differentiating between the three classes Features used from each aspect are shown in in Table \ref{tab:features_used}. The dataset used to train the classifier contains 55,301 suspended accounts, 100,000 normal accounts and 1,626 restored accounts. There is a severe class imbalance, with the minority class (restored accounts) being 1.626\% of the majority class (normal accounts).

Machine learning models are negatively affected when trained on datasets with class imbalance \cite{issue_imabalanced_1,issue_imabalanced_2}. Class imbalance refers to the issue where the number of samples from a particular class is much lower than samples from other classes. Many methods deal with class imbalance using data sampling methods. Two such methods are Random Over Sampling (ROS), which duplicates samples from the minority class, and Random Under Sampling (RUS), which eliminates samples from the majority class. Both these methods can cause issues like bias the model \cite{bias_sampling}. ROS, by oversampling may cause the model to overfit on the training data and RUS, by undersampling, may remove critical information from the dataset which is not ideal for our use case \cite{undersampling_effect}. A heuristic over-sampling algorithm, Synthetic Minority Over-Sampling Technique (SMOTE), produces artificial samples from the minority class by interpolating existing instances which are very close together. \textit{K} intraclass nearest neighbours are found for each sample in the minority class and synthetic samples are found in the direction of some or all of the nearest neighbours \cite{SMOTE}.

We used several classifier models such as Random Forest Classifier, Gradient Boosting Classifier, XGB Classifier and LGBM Classifier in combination with the above-mentioned sampling techniques. We use the best performing combination (XGB Classifier with SMOTE) for further analysis. The F1 scores for pairwise separation of the three classes are shown in Table \ref{tab:interclass_scores}.

\begin{table}
	\begin{tabular}{l}
	    \hline
		\toprule
		\textbf{Account Properties} \\
		\midrule
            Time since account creation, No. of friends, No. of followers, \\ Total no. of likes, Ratio of friends to followers, Name len in chars, \\ Bio len in chars,  Screen name len in chars, \\Screen name len in words, Bio len in words, No. of tweets, \\ Avg. no. of tweets per hour, Avg, tweet gap, Median tweet gap, \\ Avg. tweet len in words, Avg. tweets len in chars,\\  Proportion of retweets to total tweets (RT\_rate\_proportion),\\ Rate of capital chars\\
        \midrule
        \textbf{Tweet Content} \\
        \midrule
            Median Anger, Mean Anger, Median Disgust, Mean Disgust, \\ Median Fear, Mean Fear,  Median Joy, Mean Joy, \\ Median Sadness,  Mean Sadness, Median Surprise,  Mean Surprise, \\ Median Compound, Mean Compound, Median Negative, \\ Mean Negative, Median Neutral,  Mean Neutral, \\ Median Positive, Mean Positive \\
        \midrule
        \textbf{Network} \\
        \midrule
             No. of hashtags, No. of unique hashtags, No. of mentions,\\ No. of unique mentions, No. of retweets, Avg. Hashtag count, \\ Avg. unique hashtags count, Avg. mentions, Avg. unique mentions\\
		\bottomrule
	\end{tabular}
	\caption{Three categories of features (Account Properties, Tweet Content and Network) used to train the classifier.}
	\label{tab:features_used}
\end{table}

% \begin{table}
% 	\centering
% 	\begin{tabular}{lrr}
% 	    \hline
% 		\toprule
% 		\textbf{Comparison} &
% 		\textbf{F1-Score} &
% 		\textbf{Confusion Matrix} \\
% 		\midrule
%             Restored vs Normal &
%             0.71242 &
%             $\begin{bmatrix}
%             151 & 93 \\
%             4620 & 11525 
%             \end{bmatrix}$  \\ \\
%             Suspended vs Normal &
%             0.84008 &
%             $\begin{bmatrix}
%             7084 & 1656 \\
%             2338 & 13898 
%             \end{bmatrix}$  \\ \\ 
%             Suspended vs Restored &
%             0.72282 &
%             $\begin{bmatrix}
%             6383 & 2487 \\
%             48 & 228 
%             \end{bmatrix}$  \\
% 		\bottomrule
% 	\end{tabular}
% % 	\vspace{0.2cm}
% 	\caption{F1 scores and confusion matrices for pairwise prediction of account categories.}
% 	\label{tab:interclass_scores}
% \end{table}

% \begin{table}
% 	\centering
% 	\begin{tabular}{lr}
% 	    \hline
% 		\toprule
% 		\textbf{Comparison} &
% 		\textbf{F1-Score} \\
% 		\midrule
%             Restored vs Normal &
%             0.71242 \\
%             Suspended vs Normal &
%             0.84008 \\ 
%             Suspended vs Restored &
%             0.72282 \\
%         \bottomrule
% 	\end{tabular}
% % 	\vspace{0.2cm}
% 	\caption{F1 scores for pairwise prediction of account categories.}
% 	\label{tab:interclass_scores}
% \end{table}
% \vspace{-0.05cm}
\subsection{Explainability and Feature Importance}

% \subsubsection{Shapley and SHAP Values}
\label{sec:shap_values}

\begin{figure*}[ht!]
     \centering
     \begin{subfigure}[b]{0.3\textwidth}
         \centering
         \includegraphics[width=\textwidth]{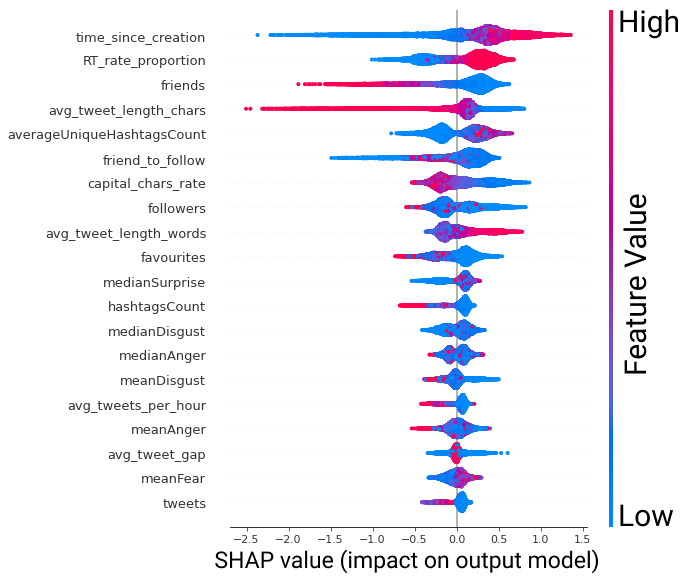}
         \caption{SHAP summary plot for the  Restored (class "0") vs Normal (class "1") accounts classifier.}\label{Fig:Restored_vs_Normal_SHAP}
     \end{subfigure}
     \hfill
     \begin{subfigure}[b]{0.29\textwidth}
         \centering
         \includegraphics[width=\textwidth]{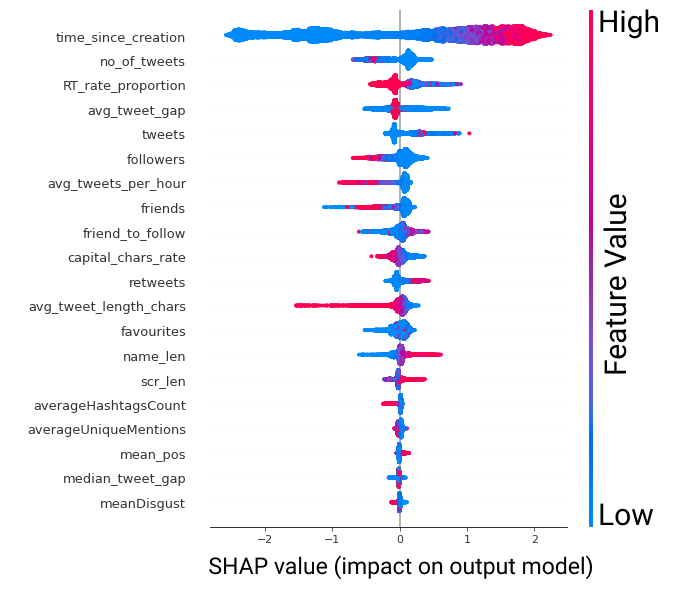}
     \caption{SHAP summary plot for the Suspended (class "0") vs Normal (class "1") accounts classifier.}\label{Fig:Suspended_vs_Normal_SHAP}
     \end{subfigure}
     \hfill
     \begin{subfigure}[b]{0.3\textwidth}
         \centering
         \includegraphics[width=\textwidth]{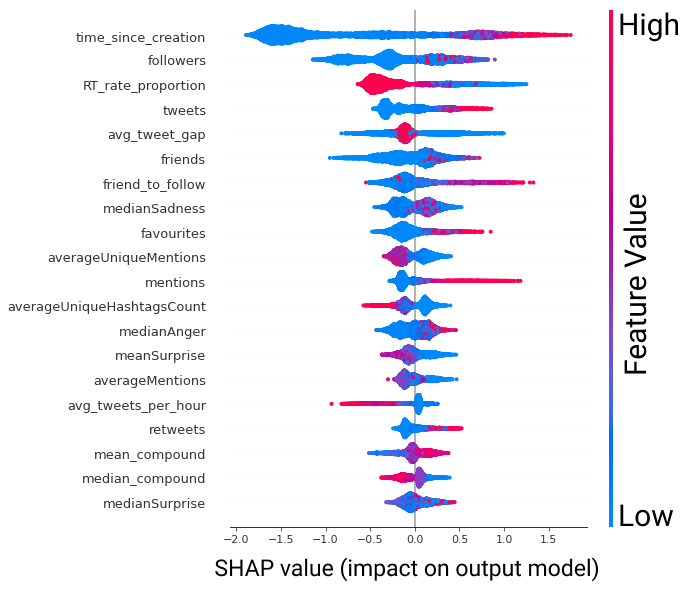}
     \caption{SHAP summary plot for the Suspended (class "0") vs Restored (class "1") accounts classifier.}
     \label{Fig:Suspended_vs_Restored_SHAP}
     \end{subfigure}
        \caption{Here, we show the SHAP summary plots generated for the three classifier models. The features in each subfigure are ranked by mean absolute value of the SHAP values for each feature. The feature value is color coded; red means that the feature value is high and blue means that the feature value is low. The X-axis gives the SHAP value for each feature value. A positive SHAP value means that the corresponding feature value is pushing the output towards class "1" and a negative SHAP value means that the corresponding feature value is pushing the output towards class "0".  }
        \label{fig:three graphs}
\end{figure*}
% \vspace{-0.1cm}
The Shapley value is a solution concept of fairly distributing gains and costs to several actors working in a coalition \cite{kuhn_17._1953}.
In the context of the explainability of machine learning models, the Shapley value allows us to identify which features contribute more to the final prediction of the model.
However, the complexity constraints of using Shapley values makes it not feasible on larger datasets.
To solve this problem, we used Shapley Additive Explanations,
or SHAP values \cite{NIPS2017_7062}, which capture the contribution of each
feature based on local explanations and principles of game theory. It uses approximate algorithms to predict the Shapley values. 
The SHAP plots in Figure \ref{fig:three graphs} represent the 20 most important features of our model sorted by the mean absolute value of the SHAP values of each feature. Here, SHAP values are used to interpret and understand which features play an important role in determining the output of each classifier.

\textit{Restored vs Normal Classifier:} When distinguishing between restored and normal accounts, the SHAP analysis from Figure \ref{Fig:Restored_vs_Normal_SHAP} shows us that higher average tweet length in characters and higher number of friends are signs of restored accounts. On the other hand, normal accounts are characterised by higher time since account creation, higher retweet rate proportion and higher average tweet length in words. Out of the top 20 features, 60\% are account property based,  30\% are tweet content based, and 10\% are network based.

\textit{Suspended vs Normal Classifier:} When distinguishing between suspended and normal accounts, the SHAP analysis from Figure \ref{Fig:Suspended_vs_Normal_SHAP} shows us that higher time since account creation, higher name length (in characters) and higher screen name length (in characters) are signs of normal accounts. When it comes to suspended accounts, a higher number of average tweets per hour and higher average tweet length (in characters) are signs of suspended accounts. The SHAP analysis also shows us that out of the top 20 features, 70\% are account property based, 20\% are network based, and 10\% are tweet content based.

\textit{Suspended vs Restored Classifier:} When distinguishing between suspended and restored accounts, the SHAP analysis from Figure \ref{Fig:Suspended_vs_Restored_SHAP} shows us that higher time since creation, higher number of mentions and higher number of tweets are signs of restored accounts. Suspended accounts have higher retweet rate proportion, higher average unique hashtags count, higher average number of tweets per hour and higher compound sentiment are signs of suspended accounts. The SHAP analysis also shows us that out of the top 20 features, 45\% are account property based, 30\% are tweet content based, and 25\% are network based.

Overall, we find that account properties often have the highest SHAP value and thus affect the model's output the most. Most notably, time since account creation and retweet rate proportion (i.e. ratio of number of retweets to tweets made by the account) have the highest impact on the outputs of the three classifiers. We find that content and network features are low in number or have very low SHAP values across all classifiers. This suggests that the reason behind suspending or restoring an account is not based on the tweet content or network properties, but rather the account properties. The lack of importance of the content features is in line with our findings in Section \ref{Content Characterstics}.

\begin{table}[H]
	\centering
	\begin{tabular}{lr}
	    \hline
		\toprule
		\textbf{Comparison} &
		\textbf{F1-Score} \\
		\midrule
            Restored vs Normal &
            0.71242 \\
            Suspended vs Normal &
            0.84008 \\ 
            Suspended vs Restored &
            0.72282 \\
        \bottomrule
	\end{tabular}
% 	\vspace{0.2cm}
	\caption{F1 scores for pairwise prediction of account categories.}
	\label{tab:interclass_scores}
\end{table}

% \vspace{-0.5cm}
\section{RQ2 - Timeline of Restored Accounts}
Aim of RQ2 was to understand what the impact of suspension was on these restored accounts. Did a ban (albeit wrong or temporary) affect the way they interacted with the platform when they came back? To critically understand the impact of the suspension event, we looked at the entire timeline of the restored accounts and split it into two phases -- pre-suspension and post-restoration. We also created a control group of accounts from the set of normal accounts to compare the change in behaviour.  
To get a better understanding, we looked at two aspects -- the amount of activity on the platform and the content produced. We study the difference in these aspects across two phases (pre-suspension and post-restoration).

\textbf{Changes in activity levels} - 
In order to attribute the change in a property of an account to suspension, there must be a measurable difference in that property pre-suspension and post-restoration.
% We measure the change in account properties between the pre-suspension to the post-restoration phase. We can then attribute any change in account property to suspension.
The properties we look at are the number of tweets posted per day per account (activity on the platform) and the different content characteristics of the tweets (Ekman emotions and VADER scores) created by the accounts. Now, to model a property, we are concerned with two aspects -- absolute value (i.e. what the current state is) and trend (i.e. moving upwards or downwards). Hence, we fit a simple straight line to any property. The y-intercept of the line gives us the absolute value while the slope of the line gives us the trend.     
% We aggregate our data on a daily basis before carrying out the above analysis.
We perform a regression discontinuity analysis  \cite{IMBENS2008615} to measure the difference in properties between the pre-suspension and post-restoration phases. We use two linear models:
\begin{equation}\label{eq:1}
y_t  = \alpha_0 + \beta_0t  \quad  (t<0)\ 
\end{equation}
% \vspace{-0.5cm}
\begin{equation}\label{eq:2}
y_t  = \alpha_1 + \beta_1t  \quad  (t>0)\ 
\end{equation}

where $t$ is the date ranging from -179 to 186 ($t$ = 0 represents June 29, 2019, i.e. the date as of which the accounts were suspended). % as shown in Section \ref{Identifying Categories of Users}).
$y_t$ represents the statistic we are modelling (tweets posted per account, Ekman and VADER scores). Equation \ref{eq:1} gives us the equation of the line pre-suspension and equation \ref{eq:2} gives us the equation of the line post-restoration. These models assume that we can approximate the various data points as a straight line defined by $\alpha_0$ and $\beta_0$ pre-suspension and $\alpha_1$ and $\beta_1$ post-restoration.

Since, we are interested in studying the change in behaviour before and after suspension, we define two additional terms: $\alpha$ and $\beta$ to represent this change. $\alpha$ is the change in the y-intercept post-restoration (ie. $\alpha  = \alpha_1 - \alpha_0$) and $\beta$ is the change in the slope post-restoration (ie. $\beta  = \beta_1 - \beta_0 $).
$\alpha$ represents the dip or rise in the absolute value caused by suspension, while $\beta$ represents the change in long-term trend. We further exclude data from a grace period before and after suspension to account for the bursty behaviour occurring in the days around suspension.

\begin{figure} [H]
  \begin{subfigure}[b]{0.45\columnwidth}
  \centering
     \includegraphics[width=\linewidth]{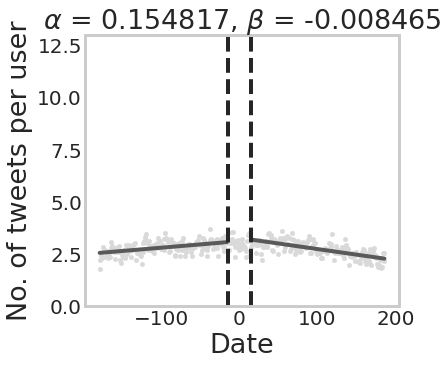}
     \caption{Normal Accounts}
     \label{Fig:regular_discontinuity}
  \end{subfigure}
  \begin{subfigure}[b]{0.45\columnwidth}
     \centering
     \includegraphics[width=\linewidth]{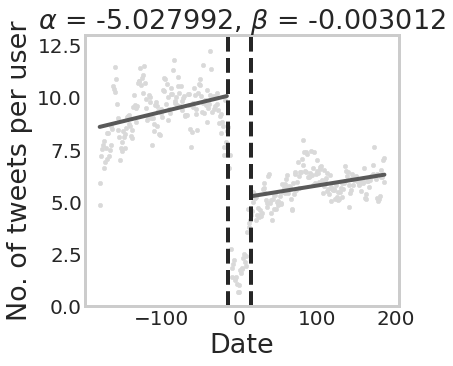}
     \caption{Restored Accounts}
     \label{Fig:restored_discontinuity}
  \end{subfigure}
\caption{Activity Levels: The daily number of tweets tweeted by (a) normal accounts and (b) restored accounts. Date "0" (29th June 2019). A grace period of 15 days before and after suspension are represented by the dotted vertical lines. On top of each subplot, we report the coefficients associated with the suspension policy ( $\alpha$ and $\beta$).}
  \label{Fig:count_discontinuity}
\end{figure}

\textbf{Amount of activity}: Figures \ref{Fig:regular_discontinuity} and \ref{Fig:restored_discontinuity} show the total number of tweets per day normalized by the number of accounts in that class before and after suspension.
% for normal and restored accounts. 
The normal accounts saw an average increase of around 0.15 tweets per day ($\alpha$ = 0.154817) after the suspension date which represents a percentage change of around +5.5\% of the \textit{M}ean \textit{V}alue \textit{B}efore  \textit{S}uspension (henceforth referred to as MVBS). On the other hand, the restored accounts saw an average decrease of about 5 tweets per day ($\alpha$ = -5.027992). This represents a percentage decrease of 53.95\% of the MVBS. This significant decrease shows that suspension does indeed have an effect on the number of tweets posted by restored accounts. We do not observe any significant change in the long-term trend ($\beta$).

\begin{figure} [H]
  \begin{subfigure}[b]{0.45\columnwidth}
  \centering
     \includegraphics[width=\linewidth]{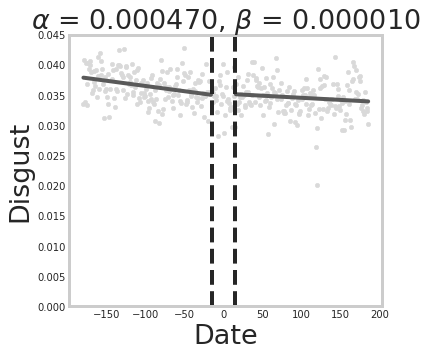}
     \caption{Ekman - Normal}
     \label{Fig:regular_discontinuity_disgust}
  \end{subfigure}
  \begin{subfigure}[b]{0.45\columnwidth}
     \centering
     \includegraphics[width=\linewidth]{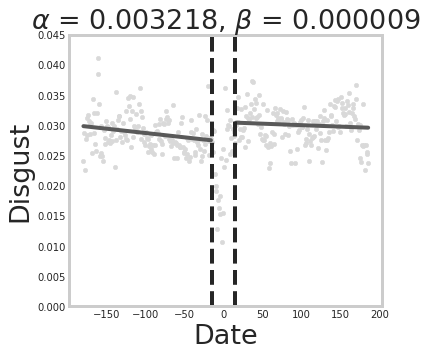}
     \caption{Ekman - Restored}
     \label{Fig:restored_discontinuity_disgust}
  \end{subfigure}
  
  \begin{subfigure}[b]{0.45\columnwidth}
  \centering
     \includegraphics[width=\linewidth]{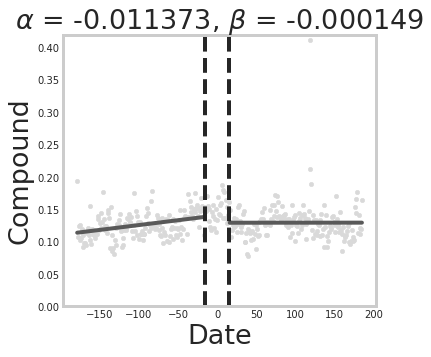}
     \caption{Vader - Normal}
     \label{Fig:regular_discontinuity_compound}
  \end{subfigure}
  \begin{subfigure}[b]{0.45\columnwidth}
     \centering
     \includegraphics[width=\linewidth]{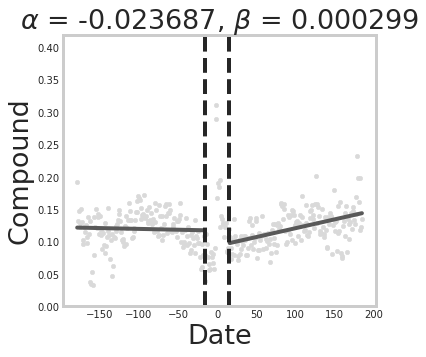}
     \caption{ Vader - Restored}
     \label{Fig:restored_discontinuity_compound}
  \end{subfigure}
  
  \caption{Content Levels: The daily Ekman (Disgust) and VADER (Compound) scores of tweets made by normal and restored accounts. The grace period and coeffecients have the same meaning as Figure \ref{Fig:count_discontinuity}. The pre-suspension and post-restoration properties for restored accounts are similar for both Ekman (b) and Vader (d).}  
  \label{Fig:discontinuity}
\end{figure}

\textbf{Content pushed}: Figures \ref{Fig:regular_discontinuity_disgust}, \ref{Fig:restored_discontinuity_disgust}, \ref{Fig:regular_discontinuity_compound} and \ref{Fig:restored_discontinuity_compound} show the average VADER and Ekman score respectively for each day before and after suspension, for normal and restored accounts. Across the six basic Ekman emotions and the four VADER sentiment classes we observe little change in behaviour. For brevity, the figure contains only Disgust (Ekman basic emotion) and Compound sentiment (VADER sentiment class), but the same was observed across all Ekman emotions and VADER sentiment scores.  

On observing the figures, we can see that is there is a negligible difference in $\alpha$ and $\beta$ values before and after the suspension for both normal and restored accounts. This is in line with our observations from section \ref{sec:shap_values} and section \ref{Content Characterstics} which show that content features do not differ much between the classes.

% \begin{figure}
% 	\centering
% 	\includegraphics[width=0.45\textwidth]{Images/count_discontinuity.png}
% 	\caption{Activity Levels: The daily number of tweets tweeted by regular users (left) and restored users (right). Date "0" (29th June 2020) is the date of suspension. A grace period of 15 days before and after suspension are represented by the dotted vertical lines. On top of each subplot, we report the coefficients associated with the suspension policy ( $\alpha$ and $\beta$).}
% 	\label{fig:count_discontinuity}
% \end{figure}

% \begin{figure*}
% 	\centering
% 	\includegraphics[width=1\textwidth]{Images/empath_discontinuity.png}
% 	\caption{Empath Score Levels: The average empath scores for tweets tweeted per day for regular users (top row) and restored users (bottom row). Date '0' (29th June 2020) is the date of suspension. A grace period of 15 days before and after suspension are represented by the dotted vertical lines. On top of each subplot, we report the coefficients associated with the suspension policy ( $\alpha$ and $\beta$).}
% 	\label{fig:empath_discontinuity}
% \end{figure*}

% \begin{figure*}
% 	\centering
% 	\includegraphics[width=1\textwidth]{Images/ekman_discontinuity.png}

% \vspace{-0.5cm}
\section{Discussion}
Restored accounts provide a new dimension to study the effect of suspension on Twitter. The most recent (9th August, 2021) case of Rep. Marjorie Taylor Greene's one-week suspension
\footnote{https://edition.cnn.com/2021/08/10/tech/twitter-marjorie-taylor-greene/index.html}, highlights the importance of studying restored accounts. We can gauge the impact of suspension by looking at changes in tweet activity and behaviour pre-suspension and post-restoration. Twitter can harness such analysis to understand the effectiveness of suspension on their platform and come up with customised suspension policies and duration of suspension to maximise effectiveness. Our work on the interpretable prediction models provides a baseline to understand the features that can impact suspension and restoration on Twitter. Restored accounts are a relatively small proportion of total accounts. To overcome this limitation we plan to scale our data collection and examine a larger more varied corpus of accounts. We also plan to expand this work to other platforms as the concept of suspension is ubiquitous across social networks.

% \vspace{-1ex}
\newcommand{\BIBdecl}{\setlength{\itemsep}{0.001 em}}
% \bibliographystyle{IEEEtran}
% argument is your BibTeX string definitions and bibliography database(s)
\bibliographystyle{ACM-Reference-Format}
\bibliography{references}

%%% -*-BibTeX-*-
%%% Do NOT edit. File created by BibTeX with style
%%% ACM-Reference-Format-Journals [18-Jan-2012].

\begin{thebibliography}{27}

%%% ====================================================================
%%% NOTE TO THE USER: you can override these defaults by providing
%%% customized versions of any of these macros before the \bibliography
%%% command.  Each of them MUST provide its own final punctuation,
%%% except for \shownote{}, \showDOI{}, and \showURL{}.  The latter two
%%% do not use final punctuation, in order to avoid confusing it with
%%% the Web address.
%%%
%%% To suppress output of a particular field, define its macro to expand
%%% to an empty string, or better, \unskip, like this:
%%%
%%% \newcommand{\showDOI}[1]{\unskip}   % LaTeX syntax
%%%
%%% \def \showDOI #1{\unskip}           % plain TeX syntax
%%%
%%% ====================================================================

\ifx \showCODEN    \undefined \def \showCODEN     #1{\unskip}     \fi
\ifx \showDOI      \undefined \def \showDOI       #1{#1}\fi
\ifx \showISBNx    \undefined \def \showISBNx     #1{\unskip}     \fi
\ifx \showISBNxiii \undefined \def \showISBNxiii  #1{\unskip}     \fi
\ifx \showISSN     \undefined \def \showISSN      #1{\unskip}     \fi
\ifx \showLCCN     \undefined \def \showLCCN      #1{\unskip}     \fi
\ifx \shownote     \undefined \def \shownote      #1{#1}          \fi
\ifx \showarticletitle \undefined \def \showarticletitle #1{#1}   \fi
\ifx \showURL      \undefined \def \showURL       {\relax}        \fi
% The following commands are used for tagged output and should be
% invisible to TeX
\providecommand\bibfield[2]{#2}
\providecommand\bibinfo[2]{#2}
\providecommand\natexlab[1]{#1}
\providecommand\showeprint[2][]{arXiv:#2}

\bibitem[\protect\citeauthoryear{Abu-El-Rub et~al\mbox{.}}{Abu-El-Rub
  et~al\mbox{.}}{2019}]%
        {abu-el-rub_botcamp:_2019}
\bibfield{author}{\bibinfo{person}{Noor Abu-El-Rub} {et~al\mbox{.}}}
  \bibinfo{year}{2019}\natexlab{}.
\newblock \showarticletitle{{BotCamp}: {Bot}-driven {Interactions} in {Social}
  {Campaigns}}. In \bibinfo{booktitle}{\emph{The {World} {Wide} {Web}
  {Conference} on - {WWW} '19}}. \bibinfo{publisher}{ACM Press},
  \bibinfo{address}{San Francisco, CA, USA}, \bibinfo{pages}{2529--2535}.
\newblock
\showISBNx{9781450366748}


\bibitem[\protect\citeauthoryear{Bowyer et~al\mbox{.}}{Bowyer
  et~al\mbox{.}}{2011}]%
        {SMOTE}
\bibfield{author}{\bibinfo{person}{Kevin~W. Bowyer} {et~al\mbox{.}}}
  \bibinfo{year}{2011}\natexlab{}.
\newblock \showarticletitle{{SMOTE:} Synthetic Minority Over-sampling
  Technique}.
\newblock \bibinfo{journal}{\emph{CoRR}}  \bibinfo{volume}{abs/1106.1813}
  (\bibinfo{year}{2011}).
\newblock
\showeprint[arxiv]{1106.1813}
\urldef\tempurl%
\url{http://arxiv.org/abs/1106.1813}
\showURL{%
\tempurl}


\bibitem[\protect\citeauthoryear{Buda et~al\mbox{.}}{Buda
  et~al\mbox{.}}{2017}]%
        {issue_imabalanced_1}
\bibfield{author}{\bibinfo{person}{Mateusz Buda} {et~al\mbox{.}}}
  \bibinfo{year}{2017}\natexlab{}.
\newblock \showarticletitle{A systematic study of the class imbalance problem
  in convolutional neural networks}.
\newblock \bibinfo{journal}{\emph{CoRR}}  \bibinfo{volume}{abs/1710.05381}
  (\bibinfo{year}{2017}).
\newblock
\showeprint[arxiv]{1710.05381}
\urldef\tempurl%
\url{http://arxiv.org/abs/1710.05381}
\showURL{%
\tempurl}


\bibitem[\protect\citeauthoryear{Chandrasekharan et~al\mbox{.}}{Chandrasekharan
  et~al\mbox{.}}{2017}]%
        {chandrasekharan_you_2017}
\bibfield{author}{\bibinfo{person}{Eshwar Chandrasekharan} {et~al\mbox{.}}}
  \bibinfo{year}{2017}\natexlab{}.
\newblock \showarticletitle{You {Can}'t {Stay} {Here}: {The} {Efficacy} of
  {Reddit}'s 2015 {Ban} {Examined} {Through} {Hate} {Speech}}.
\newblock \bibinfo{journal}{\emph{Proceedings of the ACM on Human-Computer
  Interaction}} \bibinfo{volume}{1}, \bibinfo{number}{CSCW}
  (\bibinfo{date}{Dec.} \bibinfo{year}{2017}), \bibinfo{pages}{1--22}.
\newblock
\showISSN{2573-0142}


\bibitem[\protect\citeauthoryear{Chowdhury et~al\mbox{.}}{Chowdhury
  et~al\mbox{.}}{2020}]%
        {chowdhury_twitter_2020}
\bibfield{author}{\bibinfo{person}{Farhan~Asif Chowdhury} {et~al\mbox{.}}}
  \bibinfo{year}{2020}\natexlab{}.
\newblock \showarticletitle{On {Twitter} {Purge}: {A} {Retrospective}
  {Analysis} of {Suspended} {Users}}. In \bibinfo{booktitle}{\emph{Companion
  {Proceedings} of the {Web} {Conference} 2020}}. \bibinfo{publisher}{ACM},
  \bibinfo{address}{Taipei Taiwan}, \bibinfo{pages}{371--378}.
\newblock
\showISBNx{9781450370240}


\bibitem[\protect\citeauthoryear{Colneric and Demsar}{Colneric and
  Demsar}{2020}]%
        {colneric_emotion_2020}
\bibfield{author}{\bibinfo{person}{Niko Colneric} {and} \bibinfo{person}{Janez
  Demsar}.} \bibinfo{year}{2020}\natexlab{}.
\newblock \showarticletitle{Emotion {Recognition} on {Twitter}: {Comparative}
  {Study} and {Training} a {Unison} {Model}}.
\newblock \bibinfo{journal}{\emph{IEEE Transactions on Affective Computing}}
  \bibinfo{volume}{11}, \bibinfo{number}{3} (\bibinfo{date}{July}
  \bibinfo{year}{2020}), \bibinfo{pages}{433--446}.
\newblock
\showISSN{1949-3045, 2371-9850}


\bibitem[\protect\citeauthoryear{Ekman}{Ekman}{1992}]%
        {ekman_argument_1992}
\bibfield{author}{\bibinfo{person}{Paul Ekman}.}
  \bibinfo{year}{1992}\natexlab{}.
\newblock \showarticletitle{An argument for basic emotions}.
\newblock \bibinfo{journal}{\emph{Cognition and Emotion}} \bibinfo{volume}{6},
  \bibinfo{number}{3-4} (\bibinfo{date}{May} \bibinfo{year}{1992}),
  \bibinfo{pages}{169--200}.
\newblock
\showISSN{0269-9931, 1464-0600}


\bibitem[\protect\citeauthoryear{González-Bárcenas, Rendón, Alejo,
  Granda-Gutiérrez, and Valdovinos}{González-Bárcenas et~al\mbox{.}}{2019}]%
        {bias_sampling}
\bibfield{author}{\bibinfo{person}{Victor González-Bárcenas},
  \bibinfo{person}{Erendira Rendón}, \bibinfo{person}{Roberto Alejo},
  \bibinfo{person}{Everardo Granda-Gutiérrez}, {and} \bibinfo{person}{Rosa
  Valdovinos}.} \bibinfo{year}{2019}\natexlab{}.
\newblock \bibinfo{booktitle}{\emph{Addressing the Big Data Multi-class
  Imbalance Problem with Oversampling and Deep Learning Neural Networks}}.
\newblock \bibinfo{pages}{216--224}.
\newblock
\showISBNx{978-3-030-31331-9}


\bibitem[\protect\citeauthoryear{Gupta}{Gupta}{2019}]%
        {noauthor_social_2019}
\bibfield{author}{\bibinfo{person}{Shashi Gupta}.}
  \bibinfo{year}{2019}\natexlab{}.
\newblock \bibinfo{title}{What makes Twitter so vulnerable to controversies in
  India - ET BrandEquity}.
\newblock
\newblock


\bibitem[\protect\citeauthoryear{Gupta et~al\mbox{.}}{Gupta
  et~al\mbox{.}}{2020}]%
        {singh-a}
\bibfield{author}{\bibinfo{person}{Saurabh Gupta} {et~al\mbox{.}}}
  \bibinfo{year}{2020}\natexlab{}.
\newblock \showarticletitle{\#IVoted to \#IGotPwned: Studying Voter Privacy
  Leaks in Indian Lok Sabha Elections on Twitter}.
\newblock  (\bibinfo{year}{2020}).
\newblock


\bibitem[\protect\citeauthoryear{Howard}{Howard}{2019}]%
        {annur-polisci}
\bibfield{author}{\bibinfo{person}{Jeffrey~W. Howard}.}
  \bibinfo{year}{2019}\natexlab{}.
\newblock \showarticletitle{Free Speech and Hate Speech}.
\newblock \bibinfo{journal}{\emph{Annual Review of Political Science}}
  \bibinfo{volume}{22}, \bibinfo{number}{1} (\bibinfo{year}{2019}),
  \bibinfo{pages}{93--109}.
\newblock
\showeprint{https://doi.org/10.1146/annurev-polisci-051517-012343}
\urldef\tempurl%
\url{https://doi.org/10.1146/annurev-polisci-051517-012343}
\showURL{%
\tempurl}


\bibitem[\protect\citeauthoryear{Hutto and Gilbert}{Hutto and Gilbert}{2014}]%
        {hutto_vader:_2014}
\bibfield{author}{\bibinfo{person}{C. Hutto} {and} \bibinfo{person}{Eric
  Gilbert}.} \bibinfo{year}{2014}\natexlab{}.
\newblock \showarticletitle{{VADER}: {A} {Parsimonious} {Rule}-{Based} {Model}
  for {Sentiment} {Analysis} of {Social} {Media} {Text}}.
\newblock \bibinfo{journal}{\emph{Proceedings of the International AAAI
  Conference on Web and Social Media}} \bibinfo{volume}{8}, \bibinfo{number}{1}
  (\bibinfo{date}{May} \bibinfo{year}{2014}), \bibinfo{pages}{216--225}.
\newblock
\showISSN{2334-0770}


\bibitem[\protect\citeauthoryear{Imbens and Lemieux}{Imbens and
  Lemieux}{2008}]%
        {IMBENS2008615}
\bibfield{author}{\bibinfo{person}{Guido~W. Imbens} {and}
  \bibinfo{person}{Thomas Lemieux}.} \bibinfo{year}{2008}\natexlab{}.
\newblock \showarticletitle{Regression discontinuity designs: A guide to
  practice}.
\newblock \bibinfo{journal}{\emph{Journal of Econometrics}}
  \bibinfo{volume}{142}, \bibinfo{number}{2} (\bibinfo{year}{2008}),
  \bibinfo{pages}{615--635}.
\newblock
\showISSN{0304-4076}
\newblock
\shownote{The regression discontinuity design: Theory and applications.}


\bibitem[\protect\citeauthoryear{Jungherr}{Jungherr}{2016}]%
        {twitter_literature}
\bibfield{author}{\bibinfo{person}{Andreas Jungherr}.}
  \bibinfo{year}{2016}\natexlab{}.
\newblock \showarticletitle{Twitter use in election campaigns: A systematic
  literature review}.
\newblock \bibinfo{journal}{\emph{Journal of Information Technology \&
  Politics}} \bibinfo{volume}{13}, \bibinfo{number}{1} (\bibinfo{year}{2016}),
  \bibinfo{pages}{72--91}.
\newblock
\urldef\tempurl%
\url{https://doi.org/10.1080/19331681.2015.1132401}
\showDOI{\tempurl}
\showeprint{https://doi.org/10.1080/19331681.2015.1132401}


\bibitem[\protect\citeauthoryear{Kapoor et~al\mbox{.}}{Kapoor
  et~al\mbox{.}}{2018}]%
        {kapoor_advances_2018}
\bibfield{author}{\bibinfo{person}{Kawaljeet~Kaur Kapoor} {et~al\mbox{.}}}
  \bibinfo{year}{2018}\natexlab{}.
\newblock \showarticletitle{Advances in {Social} {Media} {Research}: {Past},
  {Present} and {Future}}.
\newblock \bibinfo{journal}{\emph{Information Systems Frontiers}}
  \bibinfo{volume}{20}, \bibinfo{number}{3} (\bibinfo{date}{June}
  \bibinfo{year}{2018}), \bibinfo{pages}{531--558}.
\newblock
\showISSN{1387-3326, 1572-9419}


\bibitem[\protect\citeauthoryear{Kubat}{Kubat}{2000}]%
        {undersampling_effect}
\bibfield{author}{\bibinfo{person}{M. Kubat}.} \bibinfo{year}{2000}\natexlab{}.
\newblock \showarticletitle{Addressing the Curse of Imbalanced Training Sets:
  One-Sided Selection}.
\newblock \bibinfo{journal}{\emph{Fourteenth International Conference on
  Machine Learning}} (\bibinfo{date}{06} \bibinfo{year}{2000}).
\newblock


\bibitem[\protect\citeauthoryear{Le, Boynton, Shafiq, and Srinivasan}{Le
  et~al\mbox{.}}{2019}]%
        {le_postmortem_2019}
\bibfield{author}{\bibinfo{person}{Huyen Le}, \bibinfo{person}{G.~R. Boynton},
  \bibinfo{person}{Zubair Shafiq}, {and} \bibinfo{person}{Padmini Srinivasan}.}
  \bibinfo{year}{2019}\natexlab{}.
\newblock \showarticletitle{A Postmortem of Suspended Twitter Accounts in the
  2016 U.S. Presidential Election}. In \bibinfo{booktitle}{\emph{Proceedings of
  the 2019 IEEE/ACM International Conference on ASONAM}} (Vancouver, British
  Columbia, Canada) \emph{(\bibinfo{series}{ASONAM '19})}.
  \bibinfo{publisher}{Association for Computing Machinery},
  \bibinfo{address}{New York, NY, USA}, \bibinfo{pages}{258–265}.
\newblock
\showISBNx{9781450368681}
\urldef\tempurl%
\url{https://doi.org/10.1145/3341161.3342878}
\showDOI{\tempurl}


\bibitem[\protect\citeauthoryear{Lundberg et~al\mbox{.}}{Lundberg
  et~al\mbox{.}}{2017}]%
        {NIPS2017_7062}
\bibfield{author}{\bibinfo{person}{Scott~M Lundberg} {et~al\mbox{.}}}
  \bibinfo{year}{2017}\natexlab{}.
\newblock \showarticletitle{A Unified Approach to Interpreting Model
  Predictions}.
\newblock In \bibinfo{booktitle}{\emph{Advances in Neural Information
  Processing Systems 30}}, \bibfield{editor}{\bibinfo{person}{I.~Guyon},
  \bibinfo{person}{U.~V. Luxburg}, \bibinfo{person}{S.~Bengio},
  \bibinfo{person}{H.~Wallach}, \bibinfo{person}{R.~Fergus},
  \bibinfo{person}{S.~Vishwanathan}, {and} \bibinfo{person}{R.~Garnett}}
  (Eds.). \bibinfo{publisher}{Curran Associates, Inc.},
  \bibinfo{pages}{4765--4774}.
\newblock


\bibitem[\protect\citeauthoryear{Radicchi et~al\mbox{.}}{Radicchi
  et~al\mbox{.}}{2004}]%
        {radicchi_defining_2004}
\bibfield{author}{\bibinfo{person}{Filippo Radicchi} {et~al\mbox{.}}}
  \bibinfo{year}{2004}\natexlab{}.
\newblock \showarticletitle{Defining and identifying communities in networks}.
\newblock \bibinfo{journal}{\emph{Proceedings of the National Academy of
  Sciences}} \bibinfo{volume}{101}, \bibinfo{number}{9} (\bibinfo{date}{March}
  \bibinfo{year}{2004}), \bibinfo{pages}{2658--2663}.
\newblock


\bibitem[\protect\citeauthoryear{Ribeiro et~al\mbox{.}}{Ribeiro
  et~al\mbox{.}}{2020}]%
        {ribeiro_does_2020}
\bibfield{author}{\bibinfo{person}{Manoel~Horta Ribeiro} {et~al\mbox{.}}}
  \bibinfo{year}{2020}\natexlab{}.
\newblock \showarticletitle{Does {Platform} {Migration} {Compromise} {Content}
  {Moderation}? {Evidence} from r/{The}\_Donald and r/{Incels}}.
\newblock \bibinfo{journal}{\emph{arXiv:2010.10397 [cs]}} (\bibinfo{date}{Oct.}
  \bibinfo{year}{2020}).
\newblock
\newblock
\shownote{arXiv: 2010.10397.}


\bibitem[\protect\citeauthoryear{Shapley}{Shapley}{1953}]%
        {kuhn_17._1953}
\bibfield{author}{\bibinfo{person}{L.~S. Shapley}.}
  \bibinfo{year}{1953}\natexlab{}.
\newblock \showarticletitle{17. {A} {Value} for n-{Person} {Games}}.
\newblock In \bibinfo{booktitle}{\emph{Contributions to the {Theory} of {Games}
  ({AM}-28), {Volume} {II}}}, \bibfield{editor}{\bibinfo{person}{Harold~William
  Kuhn} {and} \bibinfo{person}{Albert~William Tucker}} (Eds.).
  \bibinfo{publisher}{Princeton University Press}, \bibinfo{pages}{307--318}.
\newblock
\showISBNx{9781400881970}


\bibitem[\protect\citeauthoryear{Thomas et~al\mbox{.}}{Thomas
  et~al\mbox{.}}{2011}]%
        {thomas_suspended_2011}
\bibfield{author}{\bibinfo{person}{Kurt Thomas} {et~al\mbox{.}}}
  \bibinfo{year}{2011}\natexlab{}.
\newblock \showarticletitle{Suspended accounts in retrospect: an analysis of
  twitter spam}. In \bibinfo{booktitle}{\emph{Proceedings of the 2011 {ACM}
  {SIGCOMM}}} \emph{(\bibinfo{series}{{IMC} '11})}. \bibinfo{address}{Berlin,
  Germany}, \bibinfo{pages}{243--258}.
\newblock
\showISBNx{9781450310130}


\bibitem[\protect\citeauthoryear{Traag et~al\mbox{.}}{Traag
  et~al\mbox{.}}{2019}]%
        {traag_louvain_2019}
\bibfield{author}{\bibinfo{person}{V.~A. Traag} {et~al\mbox{.}}}
  \bibinfo{year}{2019}\natexlab{}.
\newblock \showarticletitle{From {Louvain} to {Leiden}: guaranteeing
  well-connected communities}.
\newblock \bibinfo{journal}{\emph{Scientific Reports}} \bibinfo{volume}{9},
  \bibinfo{number}{1} (\bibinfo{year}{2019}), \bibinfo{pages}{5233}.
\newblock
\showISSN{2045-2322}


\bibitem[\protect\citeauthoryear{Volkova and Bell}{Volkova and Bell}{2017}]%
        {volkova_identifying_2017}
\bibfield{author}{\bibinfo{person}{Svitlana Volkova} {and}
  \bibinfo{person}{Eric Bell}.} \bibinfo{year}{2017}\natexlab{}.
\newblock \showarticletitle{Identifying {Effective} {Signals} to {Predict}
  {Deleted} and {Suspended} {Accounts} on {Twitter} {Across} {Languages}}.
\newblock \bibinfo{journal}{\emph{Proceedings of the International AAAI
  Conference on Web and Social Media}} \bibinfo{volume}{11},
  \bibinfo{number}{1} (\bibinfo{date}{May} \bibinfo{year}{2017}),
  \bibinfo{pages}{290--298}.
\newblock
\showISSN{2334-0770}


\bibitem[\protect\citeauthoryear{Wilkinson et~al\mbox{.}}{Wilkinson
  et~al\mbox{.}}{2016}]%
        {wilkinson_fair_2016}
\bibfield{author}{\bibinfo{person}{Mark~D. Wilkinson} {et~al\mbox{.}}}
  \bibinfo{year}{2016}\natexlab{}.
\newblock \showarticletitle{The {FAIR} {Guiding} {Principles} for scientific
  data management and stewardship}.
\newblock \bibinfo{journal}{\emph{Scientific Data}} \bibinfo{volume}{3},
  \bibinfo{number}{1} (\bibinfo{date}{Dec.} \bibinfo{year}{2016}),
  \bibinfo{pages}{160018}.
\newblock
\showISSN{2052-4463}


\bibitem[\protect\citeauthoryear{Yang et~al\mbox{.}}{Yang
  et~al\mbox{.}}{2020}]%
        {issue_imabalanced_2}
\bibfield{author}{\bibinfo{person}{Kaixiang Yang} {et~al\mbox{.}}}
  \bibinfo{year}{2020}\natexlab{}.
\newblock \showarticletitle{Hybrid Classifier Ensemble for Imbalanced Data}.
\newblock \bibinfo{journal}{\emph{IEEE Transactions on Neural Networks and
  Learning Systems}} \bibinfo{volume}{31}, \bibinfo{number}{4}
  (\bibinfo{year}{2020}), \bibinfo{pages}{1387--1400}.
\newblock
\urldef\tempurl%
\url{https://doi.org/10.1109/TNNLS.2019.2920246}
\showDOI{\tempurl}


\bibitem[\protect\citeauthoryear{Zhuravskaya et~al\mbox{.}}{Zhuravskaya
  et~al\mbox{.}}{2020}]%
        {zhuravskaya_political_2020}
\bibfield{author}{\bibinfo{person}{Ekaterina Zhuravskaya} {et~al\mbox{.}}}
  \bibinfo{year}{2020}\natexlab{}.
\newblock \showarticletitle{Political {Effects} of the {Internet} and {Social}
  {Media}}.
\newblock \bibinfo{journal}{\emph{Annual Review of Economics}}
  \bibinfo{volume}{12}, \bibinfo{number}{1} (\bibinfo{date}{Aug.}
  \bibinfo{year}{2020}), \bibinfo{pages}{415--438}.
\newblock
\showISSN{1941-1383, 1941-1391}


\end{thebibliography}
% \renewcommand{\appendixname}{Appendix~\Alph{section}}
%update 1,2,4,8
%keep 5,6,7
%add 3
%Fig.\ref{intro} with a new element labeled “xxx” to explain xxxxx
\end{document}